\documentstyle[12pt]{article}
\begin{document}
\begin{titlepage}

\title{On the construction of Fermi-Walker transported frames}
\author{J. W. Maluf$\,^{*}$and  F. F. Faria\\
Instituto de F\'{\i}sica, \\
Universidade de Bras\'{\i}lia\\
C. P. 04385 \\
70.919-970 Bras\'{\i}lia DF, Brazil\\}
\date{}
\maketitle

\begin{abstract}
We consider tetrad fields as reference frames adapted to observers 
that move along arbitrary timelike
trajectories in spacetime. By means of a local
Lorentz transformation we can transform these frames into Fermi-Walker
transported frames, which define a standard of non-rotation for
accelerated observers. Here we present a simple prescription for the
construction of Fermi-Walker transported frames out of an arbitrary 
set of tetrad fields.
\end{abstract}
\thispagestyle{empty}
\vfill
\par
\bigskip
\noindent (*) e-mail: wadih@unb.br\par
\par
\end{titlepage}
\newpage

\section{Introduction}
A large class of reference frames in spacetime experience inertial
forces, and therefore constitute noninertial frames. For instance, 
the frame adapted to an observer ``at rest" on the surface of the 
(rotating) Earth is noninertial. Observers that follow arbitrary 
timelike trajectories in spacetime will regard as natural a reference 
frame in which they are at rest, and their spatial axes do not rotate.
These observers will carry with them a set of tetrad fields such that
the timelike component is always tangent to the trajectory $C$, and 
the three spacelike components are normal to the observers'
worldlines. However, if we consider a single timelike worldline $C$,
in general the standard parallel transport of a tangent $A^\mu$ at a 
point $x^\mu$ does not lead to a tangent vector at the point 
$x^\mu + dx^\mu$ on $C$. The trajectory $C$ in general is not 
geodesic. The transport of vectors on $C$ that lead tangent vectors
into tangent vectors is realized by the Fermi-Walker transport
\cite{Synge}. The latter describes the
appropriate evolution in spacetime of noninertial frames, since the
timelike component of the tetrad field is transported into itself, 
and is always tangent to the worldline $C$ of the observer.
A Fermi-Walker transported set of tetrad fields is the best
approximation to a nonrotating reference frame in the sense of
Newtonian mechanics. It is physically realized by a system of 
gyroscopes. 

Fermi-Walker transported frames are important in several
investigations. A frame that undergoes linear and rotational 
acceleration may be described by the Frenet-Serret frame. The relative
rotational acceleration of a Frenet-Serret frame with respect to a
Fermi-Walker transported frame is taken to characterize important 
phenomena, like the gyroscopic precession \cite{Iyer}. Noninertial
reference frames in Minkowski spacetime that undergo Fermi-Walker
transport are useful, for instance, in the analysis of the inertial
effects on a Dirac particle \cite{HehlNi}. Fermi-Walker frames have 
been used in ref. \cite{CW} in the study of the geodetic and
Lense-Thirring precessions, and in the analysis of gravitational
wave resonant detectors.

A procedure for obtaining Fermi-Walker frames in the Kerr spacetime
has been worked out in ref. \cite{Marck1}. The method is based on
the property of separability of the Hamilton-Jacobi equation for the
geodesics of vacuum solutions of Petrov type D
(including the Kerr solution), and
depends on the existence of a Killing-Yano tensor that satisfies some
specific properties \cite{Marck2}.

In this paper we present a simple prescription for obtaining a
Fermi-Walker transported frame out of any set of tetrad fields, for
an arbitrary spacetime, assuming that the frame is transported along
a timelike trajectory.
The mechanism consists in finding suitable coefficients
for a local Lorentz transformation of the spatial sector of the 
tetrad field. Although the resulting equation for these coefficients
is simple, in practice it may not be straightforward to solve it in 
the general case. 

This paper is organized as follows. In section 2 we present the 
Frenet-Serret equations and the definition of the Fermi-Walker 
transport. In section 3 we recall the interpretation of tetrad fields 
as reference frames adapted to a field of observers in spacetime, and
define the acceleration tensor. This tensor determines the inertial
forces that act on the frame, and therefore may be taken to 
characterize the latter. In section 4 we show that the vanishing of
certain components of the acceleration tensor implies that the frame 
is Fermi-Walker transported. The vanishing of these components is
achieved by means of a local Lorentz transformation. We find the 
equation that the coefficients of the local Lorentz transformation
must satisfy in order to obtain the Fermi-Walker transported frame.
In section 5 we make a simple application of the results of section
4 to the determination of Fermi-Walker transported frames in 
Kerr spacetime.

Notation: spacetime indices $\mu, \nu, ...$ and 
SO(3,1) indices $a, b, ...$ run from 0 to 3. Time and space indices 
are indicated according to $\mu=0,i,\;\;a=(0),(i)$. The tetrad field 
is denoted by $e^a\,_\mu$, and the object of anholonomity reads 
$T_{a\mu\nu}=\partial_\mu e_{a\nu}-\partial_\nu e_{a\mu}$. The flat, 
Minkowski spacetime metric tensor raises and lowers tetrad indices 
and is fixed by $\eta_{ab}=e_{a\mu} e_{b\nu}g^{\mu\nu}= (-+++)$.

\section{The Frenet-Serret equations and the Fermi-Walker transport}

In this section we will adopt the notation of ref. [1] for the vector
quantities. The absolute derivative of a vector $V^\mu$ along a 
worldline $C$ is written as 

\begin{equation}
{{D V^\mu}\over {ds}}={{dV^\mu}\over {ds}}+
\Gamma^\mu_{\alpha\beta} V^\alpha{{dx^\beta}\over{ds}}\,,
\label{1}
\end{equation}
where $\Gamma^\mu_{\alpha\beta}$ are the Christoffel symbols. Let us
consider four vectors, $A^\mu$, $B^\mu$, $C^\mu$ and $D^\mu$ that
satisfy the following equations and properties:

\begin{eqnarray}
{{DA^\mu}\over {ds}}&=& b B^\mu \,, \\
{{DB^\mu}\over {ds}}&=&cC^\mu+bA^\mu \,, \\
{{DC^\mu}\over {ds}}&=&dD^\mu-cB^\mu \,, \\
{{DD^\mu}\over {ds}}&=&-dC^\mu \,, 
\label{2,3,4,5}
\end{eqnarray}
where $A^\mu A_\mu=-1$ and $B^\mu B_\mu=C^\mu C_\mu=D^\mu D_\mu=1$, 
and $b,c,d$ are nonnegative coefficients. Given the vector $A^\mu$,
eq. (2) defines $B^\mu$, eq. (3) defines $C^\mu$ and eq. (4)
defines $D^\mu$. Equation (5) is verified in view of eqs. (2) to (4).
It is not difficult to verify that eqs. (2) to (5) imply that 
$A^\mu, B^\mu, C^\mu$ and $D^\mu$ form an
orthonormal set of vectors, i.e., $A^\mu B_\mu =0$, etc.

We identify $A^\mu$ as the unit vector tangent to the trajectory $C$, 
$A^\mu=dx^\mu/ds$. In this case $A^\mu, B^\mu, C^\mu$ and $D^\mu$ 
establish the Frenet-Serret frame, and eqs. (2-5) are called the 
Frenet-Serret equations. $B^\mu$, $C^\mu$ and $D^\mu$ are the first, 
second and third normals to $C$, and $b, c, d$ are the first, second
and third curvatures of $C$, respectively \cite{Synge}. The 
Frenet-Serret orthonormal basis is suitably adapted to special curves
in spacetime. For instance, if $b=c=d=0$, the curve $C$ is a geodesic;
if $b=$ constant and $c=d=0$, $C$ represents a hyperbola, and if
$b=$ constant, $c= $ constant and $d=0$, then $C$ is a helix 
\cite{Synge}.

Let us consider a vector $F^\mu(x)$ defined on the timelike trajectory
$C$, in a spacetime determined by the metric tensor $g_{\mu\nu}$. The 
Fermi-Walker transport of $F^\mu$ on $C$ is defined by \cite{Synge}

\begin{equation}
{{D F^\mu}\over {ds}}=bF_\alpha (A^\mu B^\alpha -A^\alpha B^\mu)\,.
\label{6}
\end{equation}
Given the value $F^\mu(s_0)$ at a certain initial position $s_0$, eq. 
(6) formally determines $F^\mu$ along the curve $C$ determined by 
$x^\mu=x^\mu(s)$. The Fermi-Walker transport of a second rank tensor
along $C$ is defined by

\begin{equation}
{{D T^{\mu\nu}}\over {ds}}=
bT_\alpha\,^\nu (A^\mu B^\alpha -A^\alpha B^\mu) +
bT^\mu\,_\alpha (A^\nu B^\alpha -A^\alpha B^\nu)\,.
\label{7}
\end{equation}
It follows from the equation above that 

\begin{equation}
{{D g^{\mu\nu}}\over {ds}}= 0={{D \delta^\mu_\nu}\over {ds}}\,.
\label{8}
\end{equation}

The unit velocity 
vector $A^\mu=dx^\mu/ds$ naturally undergoes Fermi-Walker
transport. Application of eq. (6) to $A^\mu$ leads to eq. (2). It is 
also easy to show that the scalar product of two vectors is preserved
under the Fermi-Walker transport. Let $\phi$ represent the scalar
product of the vectors $\Sigma^\mu$ and $\Lambda_\mu$. Along $C$ we 
have

\begin{eqnarray}
\phi(s+ds)-\phi(s)&=&\Sigma^\mu(s+ds)\Lambda_\mu(s+ds)-
\Sigma^\mu(s)\Lambda_\mu(s) \nonumber \\
{}&=& \Lambda_\mu ( \delta^{FW}\Sigma^\mu) + 
\Sigma^\mu (\delta^{FW} \Lambda_\mu)\,,
\label{9}
\end{eqnarray}
where

\begin{eqnarray}
\delta^{FW} \Sigma^\mu&=
&-\Gamma^\mu_{\alpha\beta}\Sigma^\beta dx^\alpha+
b\Sigma_\alpha(A^\mu B^\alpha-A^\alpha B^\mu)ds \,, \nonumber \\
\delta^{FW} \Lambda_\mu&=&\Gamma^\lambda_{\alpha \mu} 
\Lambda_\lambda dx^\alpha +
b\Lambda_\alpha (A_\mu B^\alpha- A^\alpha B_\mu)ds \,.
\label{10}
\end{eqnarray}
Equations (9) and (10) imply that $\phi(s+ds)-\phi(s)=0$. 

We identify  the velocity vector on $C$ with the timelike 
component of the tetrad field $e_{(0)}\,^\mu$,

\begin{equation}
A^\mu= {{dx^\mu}\over {ds}}=e_{(0)}\,^\mu  \,.
\label{11}
\end{equation}
The Fermi-Walker transport of $e_{(0)}\,^\mu$ along $C$
guarantees that $e_{(0)}\,^\mu$ will always be tangent to $C$. The
spacelike components $e_{(k)}\,^\mu$ are everywhere orthogonal to
$e_{(0)}\,^\mu$. Along $C$, $e_{(k)}\,^\mu$ also undergoes 
Fermi-Walker transport. Since $e_{(k)}\,^\mu$ and $e_{(0)}\,^\mu$ 
are orthogonal, we have

\begin{equation}
{{D e_{(k)}\,^\mu}\over {ds}}=b e_{(0)}\,^\mu e_{(k)}\,^\lambda
B_\lambda\,.
\label{12}
\end{equation}
In view of eq. (2) the equation above may be rewritten as

\begin{equation}
{{D e_{(k)}\,^\mu}\over {ds}}= e_{(0)}\,^\mu e_{(k)\lambda}
{{De_{(0)}\,^\lambda}\over {ds}}\,.
\label{13}
\end{equation}
This equation determines the transport of the orthonormal basis
$e_a\,^\mu$ along an arbitrary timelike curve $C$, such that 
$e_{(0)}\,^\mu$ is always tangent to $C$.

\section{The tetrad field as a reference frame and the 
acceleration tensor}

In this section we will recall the discussion presented in
ref. \cite{Maluf1} regarding the characterization of tetrad fields
as reference frames in spacetime. A frame may be characterized in a
coordinate invariant way by its inertial accelerations, represented 
by the acceleration tensor. 

The notation will be slightly different from the previous section. 
The vector $A^\mu=dx^\mu/ds$ on a curve $C$ will be denoted here by 
the standard notation $u^\mu$. Thus the velocity vector of an 
observer on $C$ reads $u^\mu=dx^\mu/ds$.
We identify the observer's velocity with the $a=(0)$ 
component of $e_a\,^\mu$: $u^\mu(s)=e_{(0)}\,^\mu$.
The observer's acceleration $a^\mu$ is given by the absolute 
derivative of $u^\mu$ along $C$ \cite{Hehl3},

\begin{equation}
a^\mu= {{Du^\mu}\over{ds}} ={{De_{(0)}\,^\mu}\over {ds}} =
u^\alpha \nabla_\alpha e_{(0)}\,^\mu\,, 
\label{14}
\end{equation}
where the covariant derivative
is constructed out of the Christoffel symbols. Thus $e_a\,^\mu$
determines the velocity and acceleration along the worldline of an 
observer adapted to the frame. The set of tetrad fields for which 
$e_{(0)}\,^\mu$ describes a 
congruence of timelike curves is adapted to a class of 
observers characterized by the velocity field 
$u^\mu=e_{(0)}\,^\mu$ and by the acceleration $a^\mu$. 

We may consider not only the acceleration of observers along
trajectories whose tangent vectors are given by $e_{(0)}\,^\mu$, but
the acceleration of the whole frame along $C$. The 
acceleration of the frame is determined by the absolute derivative
of $e_a\,^\mu$ along the path $x^\mu(s)$. Thus, assuming that the 
observer carries an orthonormal tetrad frame $e_a\,^\mu$, the 
acceleration of the latter along the path is given by 
\cite{Mashh2}

\begin{equation}
{{D e_a\,^\mu} \over {ds}}=\phi_a\,^b\,e_b\,^\mu\,,
\label{15}
\end{equation}
where $\phi_{ab}$ is the antisymmetric acceleration tensor. 
According to ref. \cite{Mashh2}, 
in analogy with the Faraday tensor we can identify
$\phi_{ab} \rightarrow ({\bf a}, {\bf \Omega})$, where 
${\bf a}$ is the translational acceleration ($\phi_{(0)(i)}=a_{(i)}$)
and ${\bf \Omega}$ is the angular velocity 
of the local spatial frame  with respect to a nonrotating
(Fermi-Walker transported) frame. 
It follows from Eq. (11) that

\begin{equation}
\phi_a\,^b= e^b\,_\mu {{D e_a\,^\mu} \over {ds}}=
e^b\,_\mu \,u^\lambda\nabla_\lambda e_a\,^\mu\,.
\label{16}
\end{equation}

Therefore given any set of tetrad fields for an arbitrary 
gravitational field configuration, its geometrical interpretation
may be obtained by suitably interpreting the velocity field 
$u^\mu=\,e_{(0)}\,^\mu$ and the acceleration tensor $\phi_{ab}$.
The acceleration vector $a^\mu$ defined by Eq. (14) 
may be projected on a frame in order to yield

\begin{equation}
a^b= e^b\,_\mu a^\mu=e^b\,_\mu u^\alpha \nabla_\alpha
e_{(0)}\,^\mu=\phi_{(0)}\,^b\,.
\label{17}
\end{equation}
Thus $a^\mu$ and $\phi_{(0)(i)}$ are not different 
accelerations of the frame. 

The acceleration $a^\mu$ given by Eq. (14) may be rewritten as

\begin{eqnarray}
a^\mu&=& u^\alpha \nabla_\alpha e_{(0)}\,^\mu 
=u^\alpha \nabla_\alpha u^\mu =
{{dx^\alpha}\over {ds}}\biggl(
{{\partial u^\mu}\over{\partial x^\alpha}}
+\Gamma^\mu_{\alpha\beta}u^\beta \biggr) \nonumber \\
&=&{{d^2 x^\mu}\over {ds^2}}+\Gamma^\mu_{\alpha\beta}
{{dx^\alpha}\over{ds}} {{dx^\beta}\over{ds}}\,,
\label{18}
\end{eqnarray}
where $\Gamma^\mu_{\alpha\beta}$ are the Christoffel symbols.
Thus if $u^\mu=e_{(0)}\,^\mu$ represents a geodesic
trajectory, then the frame is in free fall and 
$a^\mu=0=\phi_{(0)(i)}$. Therefore we conclude that nonvanishing
values of $\phi_{(0)(i)}$ represent inertial accelerations
of the frame.

Following ref. \cite{Maluf1}, we take into account the 
orthogonality of the tetrads and write Eq. (16) as
$\phi_a\,^b= -u^\lambda e_a\,^\mu \nabla_\lambda e^b\,_\mu$, 
where $\nabla_\lambda e^b\,_\mu=\partial_\lambda e^b\,_\mu-
\Gamma^\sigma_{\lambda \mu} e^b\,_\sigma$. Now we consider
the identity $\partial_\lambda e^b\,_\mu-
\Gamma^\sigma_{\lambda \mu} e^b\,_\sigma+\,\,
^0\omega_\lambda\,^b\,_c e^c\,_\mu=0$, where
$^0\omega_\lambda\,^b\,_c$ is the metric compatible, torsion-free
Levi-Civita connection, and express $\phi_a\,^b$ according to

\begin{equation}
\phi_a\,^b=e_{(0)}\,^\mu(\,\,^0\omega_\mu\,^b\,_a)\,.
\label{19}
\end{equation}
Finally we take into account the identity 
$\,\,^0\omega_\mu\,^a\,_b= -K_\mu\,^a\,_b$, where $-K_\mu\,^a\,_b$ 
are the Ricci rotation coefficients defined by

\begin{equation}
K_{\mu ab}={1\over 2}e_a\,^\lambda e_b\,^\nu(T_{\lambda \mu\nu}+
T_{\nu\lambda\mu}+T_{\mu\lambda\nu})\,,
\label{20}
\end{equation}
and $T_{\lambda \mu\nu}=e^a\,_\lambda T_{a\mu\nu}$. 
After simple manipulations we arrive at

\begin{equation}
\phi_{ab}={1\over 2} \lbrack T_{(0)ab}+T_{a(0)b}-T_{b(0)a}
\rbrack\,.
\label{21}
\end{equation}

The expression above is not invariant under local SO(3,1) 
transformations, and for this reason the values of $\phi_{ab}$
may characterize the frame.
However, eq. (21) is invariant under coordinate transformations. We 
interpret $\phi_{ab}$ as the inertial accelerations along the 
trajectory $C$.

\section{Construction of Fermi-Walker frames}

Let us consider the expression for the Fermi-Walker transport of
tetrad fields given by eq. (13). Taking into account eq. (14) we may
rewrite (13) as

\begin{equation}
{{D e_{(k)}\,^\mu}\over {ds}}= u^\mu e_{(k)\lambda}a^\lambda
\,.
\label{22}
\end{equation}
In view of (17) we have 
$e_{(k)\lambda}a^\lambda=a_{(k)}=\phi_{(0)(k)}$. Therefore the 
Fermi-Walker transport of a frame may be written as

\begin{equation}
{{D e_{(k)}\,^\mu}\over {ds}}= u^\mu \phi_{(0)(k)}\,.
\label{23}
\end{equation}

On the other hand, it is easy to verify that the total acceleration
of the frame components $e_{(k)}\,^\mu$ given by eq. (15) may be
expressed in terms of $\phi_{(0)(k)}$ and $\phi_{(j)(k)}$ as 

\begin{equation}
{{D e_{(k)}\,^\mu}\over {ds}}= u^\mu \phi_{(0)(k)}
+\phi_{(k)}\,^{(j)} e_{(j)}\,^\mu\,.
\label{24}
\end{equation}
Therefore if $\phi_{(j)(k)}=0$, the frame is Fermi-Walker transported,
in agreement with the discussion after eq. (15). It turns out that we
can easily require $\phi_{(j)(k)}=0$, at least formally.

The expression of $\phi_{(i)(j)}$ is given by
\begin{equation}
\phi_{(i)(j)}={1\over 2} \lbrack
e_{(i)}\,^\mu e_{(j)}\,^\nu T_{(0)\mu\nu} +
e_{(0)}\,^\mu e_{(j)}\,^\nu T_{(i)\mu\nu} -
e_{(0)}\,^\mu e_{(i)}\,^\nu T_{(j)\mu\nu}\rbrack\,.
\label{25}
\end{equation}
We perform a local Lorentz rotation,

\begin{equation}
\tilde{e}_{(i)}\,^\mu = \Lambda_{(i)}\,^{(k)} e_{(k)}\,^\mu \,,
\label{26}
\end{equation}
that yields

\begin{eqnarray}
\tilde{T}_{(i)\mu\nu}&=&
\partial_\mu \tilde{e}_{(i)\nu} -
\partial_\nu \tilde{e}_{(i)\mu} \nonumber \\
{}&=&\Lambda_{(i)}\,^{(k)} T_{(k)\mu\nu}+
\lbrack \partial_\mu \Lambda_{(i)}\,^{(k)}\rbrack e_{(k)\nu}-
\lbrack \partial_\nu \Lambda_{(i)}\,^{(k)}\rbrack e_{(k)\mu}\,.
\label{27}
\end{eqnarray}
The local Lorentz coefficients  
$\lbrace \Lambda_{(i)}\,^{(j)} \rbrace$
will be fixed such that $\tilde{\phi}_{(i)(j)}=0$.

Equations (26) and (27) imply

\begin{eqnarray}
\tilde{\phi}_{(i)(j)}&=& 
\Lambda_{(i)}\,^{(k)} \Lambda_{(j)}\,^{(l)} \phi_{(k)(l)}\nonumber \\
&{}&
+{1\over 2}\biggl\{
e_{(0)}\,^\mu \Lambda_{(j)}\,^{(k)} e_{(k)}\,^\nu
\biggl[\lbrack \partial_\mu \Lambda_{(i)}\,^{(l)} \rbrack
e_{(l)\nu} -
\lbrack \partial_\nu \Lambda_{(i)}\,^{(l)}\rbrack e_{(l)\mu}\biggr]
\nonumber \\
&&-
e_{(0)}\,^\mu \Lambda_{(i)}\,^{(k)} e_{(k)}\,^\nu
\biggl[\lbrack \partial_\mu \Lambda_{(j)}\,^{(l)} \rbrack
e_{(l)\nu} -
\lbrack \partial_\nu \Lambda_{(j)}\,^{(l)}\rbrack e_{(l)\mu}\biggr]
\biggr\} \nonumber \\
&=&
\Lambda_{(i)}\,^{(k)} \Lambda_{(j)}\,^{(l)} \phi_{(k)(l)}\nonumber \\
&&
+{1\over 2}\biggl\{
e_{(0)}\,^\mu \Lambda_{(j)(k)}
\lbrack \partial_\mu \Lambda_{(i)}\,^{(k)}\rbrack -
e_{(0)}\,^\mu \Lambda_{(i)(k)}\lbrack
\partial_\mu \Lambda_{(j)}\,^{(k)}\rbrack
\biggr\}\,.
\label{28}
\end{eqnarray}
The equation above may be written in a more convenient way as

\begin{eqnarray}
\tilde{\phi}_{(i)(j)}&=& {1\over 2} \biggl[
\Lambda_{(j)}\,^{(m)} \biggl(
\Lambda_{(i)}\,^{(k)} \phi_{(k)(m)}+
e_{(0)}\,^\mu \partial_\mu \Lambda_{(i)(m)}\biggr) \nonumber \\
&& -
\Lambda_{(i)}\,^{(m)} \biggl(
\Lambda_{(j)}\,^{(k)} \phi_{(k)(m)}+
e_{(0)}\,^\mu \partial_\mu \Lambda_{(j)(m)}\biggr)
\biggr]\,.
\label{29}
\end{eqnarray}

Therefore we obtain $\tilde{\phi}_{(i)(j)}=0$ if

\begin{equation}
\Lambda_{(i)}\,^{(k)} \phi_{(k)(m)}+
e_{(0)}\,^\mu \partial_\mu \Lambda_{(i)(m)}\,,
\label{30}
\end{equation}
or, equivalently

\begin{equation}
e_{(0)}\,^\mu \Lambda^{(j)}\,_{(m)} \partial_\mu \Lambda_{(j)(k)}-
\phi_{(k)(m)}=0\,.
\label{31}
\end{equation}

The equation above is the main result of the paper:
given an arbitrary frame transported along $e_{(0)}\,^\mu$,
we calculate the angular velocity
$\phi_{(k)(m)}$, and by means of eq. (31) we determine 
the coefficients $\lbrace \Lambda_{(i)}\,^{(j)} \rbrace$. Equation
(31) ensures that the frame obtained according to eq. (26)
is Fermi-Walker transported. We note that the local Lorentz rotation
does not affect the timelike component $e_{(0)}\,^\mu$.

For an arbitrary set of tetrad fields it may not be straightfoward to
solve eq. (31) for $\lbrace \Lambda_{(i)}\,^{(j)} \rbrace$. However, 
in the weak field approximation, or in case the spacetime is 
asymptotically flat and we have

\begin{equation}
e^a\,_\mu\cong \delta^a_\mu + {1\over 2} h^a\,_\mu \,,
\label{32}
\end{equation}
we may find the approximate expression of $\Lambda_{(i)}\,^{(j)}$.
In this case we may write

\begin{equation}
\Lambda_{(j)(k)}\cong \eta_{(j)(k)}+\varepsilon_{(j)(k)}\,.
\label{33}
\end{equation}
We assume that both $h^a\,_\mu$ and $\varepsilon_{(j)(k)}$ are of 
order $\epsilon$, where $\epsilon <<1$. Then $\phi_{(j)(k)}$
is also of order $\epsilon$. In this approximation we have
$e_{(0)}\,^\mu \partial_\mu \cong \partial_0=\partial/\partial t$. 
Under these conditions we may solve eq. (31) and obtain

\begin{equation}
\Lambda_{(j)(k)}=
\pmatrix{1&\varepsilon_{(1)(2)}&\varepsilon_{(1)(3)}\cr
-\varepsilon_{(1)(2)}&1&\varepsilon_{(2)(3)}\cr
-\varepsilon_{(1)(3)}&-\varepsilon_{(2)(3)}&1\cr}\,,
\label{34}
\end{equation}
where

\begin{equation}
\dot{\varepsilon}_{(j)(k)}=\phi_{(k)(j)}\,.
\label{35}
\end{equation}
The dot denotes time derivative.

\section{Fermi-Walker frames in the Kerr spacetime}

Although the solution of eq. (31) in the general case (for an
arbitrary frame) is not always feasible, in certain situations we
obtain simple and interesting results. We will consider here the 
frame addressed in ref. \cite{Maluf1} that describes static 
observers in the Kerr spacetime, and construct Fermi-Walker
transported frames in the equatorial plane $\theta=\pi/2$.

The Kerr spacetime is determined by the line element

\begin{eqnarray}
ds^2&=&
-{{\psi^2}\over {\rho^2}}dt^2-{{2\chi\sin^2\theta}\over{\rho^2}}
\,d\phi\,dt
+{{\rho^2}\over {\Delta}}dr^2 \nonumber \\
&{}&+\rho^2d\theta^2+ {{\Sigma^2\sin^2\theta}\over{\rho^2}}d\phi^2\,,
\label{36}
\end{eqnarray}
with the following definitions:

\begin{eqnarray}
\Delta&=& r^2+a^2-2mr\,,  \nonumber \\
\rho^2&=& r^2+a^2\cos^2\theta \,, \nonumber \\
\Sigma^2&=&(r^2+a^2)^2-\Delta a^2\sin^2\theta\,,  \nonumber \\
\psi^2&=&\Delta - a^2 \sin^2\theta\,,  \nonumber \\
\chi &=&2amr\,.
\label{37}
\end{eqnarray}

A static reference frame in Kerr's spacetime is defined by the
congruence of timelike curves $u^\mu(s)$ such that $u^i=0$,
i.e., the spatial velocity of the observers is zero with 
respect to static observers at spacelike infinity. Since we
identify $u^i=e_{(0)}\,^i$, a static reference frame is
established by the condition

\begin{equation}
e_{(0)}\,^i=0\,.
\label{38}
\end{equation}
In view of the orthogonality of the tetrads, the equation above
implies $e^{(k)}\,_0=0$. This latter equation remains satisfied
even after a local rotation of the frame, 
$\tilde e^{(k)}\,_0=\Lambda^{(k)}\,_{(j)} e^{(j)}\,_0=0$.
Therefore condition (38) determines the static character of the
frame, up to an orientation of the frame in the three-dimensional
space. 

A simple form for the tetrad field that satisfies Eq. (38) reads 

\begin{equation}
e_{a\mu}=\pmatrix{-{\cal A}&0&0&-{\cal B}\cr
0&{\cal C}\sin\theta\cos\phi& \rho\cos\theta\cos\phi&-
{\cal D}\sin\theta\sin\phi\cr
0&{\cal C}\sin\theta\sin\phi& \rho\cos\theta\sin\phi&
{\cal D}\sin\theta\cos\phi\cr
0&{\cal C}\cos\theta&-\rho\sin\theta&0}\,,
\label{39}
\end{equation}
with the definitions

\begin{eqnarray}
{\cal A}&=& {\psi \over \rho}\,,  \nonumber \\
{\cal B}&=& {{\chi \sin^2\theta}\over {\rho \psi}}\,,  \nonumber \\
{\cal C}&=&{\rho \over \sqrt{\Delta}}\,, \nonumber \\
{\cal D}&=& {\Lambda \over{\rho \psi}}\,,
\label{40}
\end{eqnarray}
and $\Lambda =(\psi^2\Sigma^2+\chi^2\sin^2\theta)^{1/2}$
($a$ and $\mu$ represent lines and columns, respectively).
The frame (39) is additionally fixed by the condition that the
vector $e_{(3)}\,^\mu$ is oriented along the $z$ 
axis \cite{Maluf1}.

The linear acceleration and angular velocity that are necessary
to cancel the gravitational forces on the frame, and maintain it 
static in spacetime, are given by $\phi_{ab}$. The 
acceleration tensor was calculated in ref. \cite{Maluf1}. We define

\begin{eqnarray}
{\bf a}&=& (\phi_{01}, \phi_{02}, \phi_{03})\,, \\
{\bf \Omega}&=& (\phi_{23}, \phi_{31}, \phi_{12})\,.
\label{41,42}
\end{eqnarray}
The expressions for ${\bf a}$ and ${\bf \Omega}$ are given by

\begin{eqnarray}
{\bf a}&=& {m \over \psi^2}\biggl[
{{\sqrt{\Delta} \over {\rho}}\biggl(
{{2r^2}\over \rho^2}-1 \biggr){\bf \hat{r}}\, +
{{2ra^2}\over {\rho}^3}}\sin\theta \cos\theta\,{\bf \hat{\theta}}
\biggr]\,,\\
{\bf \Omega}&=& -{\chi \over{\Lambda \rho }}\cos\theta\,{\bf \hat{r}}+
{{\psi^2 \sqrt{\Delta}}\over {2\Lambda \rho}}\sin\theta\,
\partial_r\biggl( {\chi \over {\psi^2}}\biggr)\, {\bf \hat{\theta}}
-{{\psi^2} \over {2\Lambda \rho}}\sin\theta\,\partial_\theta 
\biggl({\chi \over {\psi^2}} \biggr) {\bf \hat{r}},
\label{43,44}
\end{eqnarray}
where

\begin{eqnarray}
{\bf \hat{r}}&=& \sin\theta \cos\phi\,{\bf \hat{x}}+
\sin\theta \sin\phi\,{\bf \hat{y}}+
\cos\theta \,{\bf \hat{z}}, \nonumber \\
{\bf \hat{\theta}}&=&\cos\theta \cos\phi\,{\bf \hat{x}}+
\cos\theta \sin\phi\,{\bf \hat{y}}-
\sin\theta\, {\bf \hat{z}}.
\label{45}
\end{eqnarray}

Restricting the analysis to the plane $\theta=\pi/2$, it is easy to
verify that ${\bf \Omega}$ reduces to

\begin{equation}
{\bf \Omega} = \phi_{(1)(2)} {\bf \hat{z}}=
-{{\psi^2 \sqrt{\Delta}}\over {2\Lambda \rho}}
\partial_r\biggl({\chi\over \psi^2}\biggr) {\bf \hat{z}}\,.
\label{46}
\end{equation}

Returning now to eq. (31), we observe that 
the general structure of the matrix $\Lambda_{(i)(j)}$ may be given
in terms of Euler angles. There are several possible ways of writing 
a rotation matrix as function of Euler angles. For our purposes,
a convenient way is given in ref. \cite{HG} (appendix A, eq. (A.3y)).
It reads

$$\Lambda_{(i)(j)}=$$

\begin{equation}
\pmatrix{
-\sin\alpha\,\sin\beta+\cos\alpha\,\cos\beta\,\cos\gamma&
\cos\alpha\,\sin\beta+\sin\alpha\,\cos\beta\,\cos\gamma&
-\cos\beta\,\sin\gamma\cr
-\sin\alpha\,\cos\beta-\cos\alpha\,\sin\beta\,\cos\gamma&
\cos\alpha\,\cos\beta-\sin\alpha\,\sin\beta\,\cos\gamma&
\sin\beta\,\sin\gamma\cr
\cos\alpha\,\sin\gamma&
\sin\alpha\,\sin\gamma&
\cos\gamma}.
\label{47}
\end{equation}
We assume that $\alpha, \beta$ and $\gamma$ are functions (to be 
determined) of the time parameter $t$.

The reason for choosing the expression above is the following. In eq.
(31) we have $e_{(0)}\,^\mu \partial_\mu =e_{(0)}\,^0 \partial_0$. 
We evaluate the quantities below,

\begin{eqnarray}
\Lambda^{(k)}\,_{(3)}\partial_0\Lambda_{(k)(2)}&=&
-\dot{\gamma}\sin\alpha+\dot{\beta}\cos\alpha\,\sin\gamma
=-\omega_x\,,
\nonumber \\
\Lambda^{(k)}\,_{(1)}\partial_0\Lambda_{(k)(3)}&=&
-\dot{\gamma}\cos\alpha-\dot{\beta}\sin\alpha\,\sin\gamma
=-\omega_y\,,
\nonumber \\
\Lambda^{(k)}\,_{(2)}\partial_0\Lambda_{(k)(1)}&=&
-\dot{\beta}\cos\gamma - \dot{\alpha}=-\omega_z\,,
\label{48}
\end{eqnarray}
where the dot represents time derivative, and note that the right 
hand side of the expressions in eq. (48) are, except for the sign,
the angular velocities along the space axes ($\omega_x, \omega_y$
and $\omega_z$ in eq. (A.8y) of ref. \cite{HG}). 

The general solution of eq. (31)
amounts to determining $\alpha, \beta$ and $\gamma$ where

\begin{eqnarray}
e_{(0)}\,^0(
-\dot{\gamma}\sin\alpha+\dot{\beta}\cos\alpha\,\sin\gamma)&=&
\phi_{(2)(3)}\,, \nonumber \\
e_{(0)}\,^0(
-\dot{\gamma}\cos\alpha-\dot{\beta}\sin\alpha\,\sin\gamma)&=&
\phi_{(3)(1)}\,, \nonumber \\
e_{(0)}\,^0(
-\dot{\beta}\cos\gamma - \dot{\alpha})&=&
\phi_{(1)(2)}\,.
\label{49}
\end{eqnarray}
Clearly there is no simple solution to $\alpha, \beta$ and $\gamma$.
However, if we restrict the analysis to the equatorial plane 
defined by $\theta=\pi/2$, the problem is greatly simplified. We 
recall that in the equatiorial plane ${\bf \Omega}$ given by eq. (46)
is directed along the $z$ axis. Thus eq. (47) must describe a 
rotation along the $z$ axis. Making $\beta=\gamma=0$ we have

\begin{equation}
\Lambda_{(i)(j)}=\pmatrix{
\cos\alpha&\sin\alpha&0\cr
-\sin\alpha&\cos\alpha&0\cr
0&0&1}\,,
\label{50}
\end{equation}
and as a consequence,

\begin{equation}
\Lambda^{(k)}\,_{(2)}\partial_0\Lambda_{(k)(1)}=-\dot{\alpha}\,.
\label{51}
\end{equation}
Taking into account eq. (46) we write

\begin{equation}
- e_{(0)}\,^0\;\dot{\alpha}=\phi_{(1)(2)}=
-{{\psi^2 \sqrt{\Delta}}\over {2\Lambda \rho}}
\partial_r\biggl({\chi\over \psi^2}\biggr)\,,
\label{52}
\end{equation}
from what we obtain

\begin{equation}
\alpha= {{\psi^3 \sqrt{\Delta}}\over{2\rho^2 \Lambda}}\partial_r
\biggl({\chi \over \psi^2}\biggr) t\,.
\label{53}
\end{equation}

Equation (50) acts only on the spatial sector of the tetrad field, 
and allows obtaining the transformed tetrads according to eq. (26).
Dropping the tilde, the transformed tetrad field 
$e_{a\mu}(t,r,\theta =\pi/2, \phi)$ is given by

\begin{eqnarray}
e_{(0)\mu}&=&(-{\psi \over \rho},0,0,
-{\chi \over{\rho \psi}}) \nonumber \\
e_{(1)\mu}&=&(0,
{\rho \over\sqrt{\Delta}}(\cos\phi\,\cos\alpha+\sin\phi\,\sin\alpha)
\,,0\,,
{\Lambda\over{\rho\psi}}(-\sin\phi\,\cos\alpha+\cos\phi\,\sin\alpha))
\nonumber \\
e_{(2)\mu}&=&(0,
{\rho \over\sqrt{\Delta}}(-\cos\phi\,\sin\alpha+\sin\phi\,\cos\alpha)
\,,0\,,
{\Lambda\over{\rho\psi}}(\sin\phi\,\sin\alpha+\cos\phi\,\cos\alpha))
\nonumber \\
e_{(3)\mu}&=&(0,0,-\rho\,,0)\,.
\label{54}
\end{eqnarray}

The frame above is Fermi-Walker transported, and is adapted to 
observers located at the equatorial plane. It is possible to check
by direct calculations that $\phi_{(i)(j)}=0$ if the latter is 
evaluated out of the tetrads above.

Another simple construction of Fermi-Walker transported tetrad fields
is the frame for observers located on the $z$ axis, namely, for
$\theta=0$. Equation (44) reduces to

\begin{equation}
{\bf \Omega}=-{\chi \over{\Lambda \rho}} {\bf \hat{z}}\,.
\label{55}
\end{equation}
The rotation matrix is again given by eq. (50). Similar to eq.
(52) we find

\begin{equation}
- e_{(0)}\,^0\;\dot{\alpha}=\phi_{(1)(2)}=
-{\chi \over{\Lambda \rho}}\,,
\label{56}
\end{equation}
and thus we obtain

\begin{equation}
\alpha= {{\chi \psi}\over {\Lambda \rho^2}} t\,.
\label{57}
\end{equation}

We remark that the frame presented in ref. \cite{Marck1} for the
equatorial plane of the Kerr spacetime does not have a simple form
as eq. (54), and is not adapted to static observers (as eq. (39))
because it satisfies $e^{(1)}\,_0 \ne 0 \ne e^{(3)}\,_0$, and
consequently $e_{(0)}\,^i \ne 0$.

\section{Conclusions}

We have found a mechanism for constructing Fermi-Walker 
frames out of an arbitrary frame transported along a curve $C$,
whose tangent vector is given by the observer's velocity
$e_{(0)}\,^\mu=dx^\mu /ds$.
The idea consists in performing a local Lorentz rotation and solving
eq. (31) for the coefficients of the local Lorentz transformation.
Although in principle it is difficult to obtain the general 
solution of eq. (31), in some particular situations the solution can 
be easily achieved. We have obtained Fermi-Walker transported frames
for observers located in the equatorial plane and in the symmetry
axis (the $z$ axis) in
the Kerr spacetime. It is formally possible to establish Fermi-Walker
frames for observers at rest on the surface of the 
rotating Earth (the Frenet-Serret equations would determine the
trajectory of the observer in spacetime). 
These frames are the best realization of reference
frames that take into account the effects of the rotation of the
Earth.

\bigskip 
\noindent {\bf Acknowledgement}\par
\noindent This work was supported in part by CNPQ (Brazil).


\begin{thebibliography}{99}

\bibitem{Synge}
J. L. Synge, {\it Relativity: The General Theory} 
(North Holland, Amsterdam, 1960).

\bibitem{Iyer}
B. R. Iyer and C. V. Vishveshwara, Phys. Rev. D {\bf 48} (1993) 5706.

\bibitem{HehlNi}
F. W. Hehl and W. -T. Ni, Phys. Rev. D {\bf 42} (1990) 2045.

\bibitem{CW}
I. Ciufolini and J. A. Wheeler, {\it Gravitation and Inertia}
(Princeton Univ. Press, Princeton, 1995).

\bibitem{Marck1}
J. A. Marck, Proc. R. Soc. Lond. A {\bf 385} (1983) 431; Phys.
Lett. A {\bf 97} (1983) 140.

\bibitem{Marck2}
N. Kamran and J. A. Marck, J. Math. Phys. {\bf 27} (1986) 1589.

\bibitem{Maluf1}
J. W. Maluf, F. F. Faria and S. C. Ulhoa, Class. Quantum Grav.
{\bf 24}, (2007) 2743-2753 [arXiv:0704.0986].

\bibitem{Hehl3}
F. H. Hehl, J. Lemke and E. W. Mielke, 
``Two Lectures on Fermions and Gravity", in {\it Geometry 
and Theoretical Physics}, edited by J. Debrus and A. C. 
Hirshfeld (Springer, Berlin Heidelberg, 1991).

\bibitem{Mashh2}
B. Mashhoon and U. Muench, Ann. Phys. (Leipzig) {\bf 11} (2002) 532
[gr-qc/0206082].

\bibitem{HG}
H. Goldstein, C. Poole and J. Safko, {\it Classical Mechanics},
third edition (Addison-Wesley, San Francisco, 2001).

\end{thebibliography}
\end{document}